# Assessing Vaccination Sentiments with Online Social Media: Implications for Infectious Disease Dynamics and Control

**Marcel Salathé\* & Shashank Khandelwal**

*Center for Infectious Disease Dynamics, Department of Biology, Penn State University*

**\*Corresponding author:**

Marcel Salathé, Department of Biology, Penn State University

Mueller 516, University Park, PA 18062, USA

Tel: (814) 867-4431

E-mail: salathe@psu.edu

# ABSTRACT

There is great interest in the dynamics of health behaviors in social networks and how they affect collective public health outcomes, but measuring population health behaviors over time and space requires substantial resources. Here, we use publicly available data from 101,853 users of online social media collected over a time period of almost six months to measure the spatio-temporal sentiment towards a new vaccine. We validated our approach by identifying a strong correlation between sentiments expressed online and CDC-estimated vaccination rates by region. Analysis of the network of opinionated users showed that information flows more often between users who share the same sentiments - and less often between users who do not share the same sentiments - than expected by chance alone. We also found that most communities are dominated by either positive or negative sentiments towards the novel vaccine. Simulations of infectious disease transmission show that if clusters of negative vaccine sentiments lead to clusters of unprotected individuals, the likelihood of disease outbreaks are greatly increased. Online social media provide unprecedented access to data allowing for inexpensive and efficient tools to identify target areas for intervention efforts and to evaluate their effectiveness.

# AUTHOR SUMMARY

Sentiments about vaccination can strongly affect individual vaccination decisions. Measuring such sentiments - and how they are distributed in a population - is typically a difficult and resource-intensive endeavor. We use publicly available data from Twitter, a popular online social media service, to measure the evolution and distribution of sentiments towards the novel influenza A(H1N1) vaccine during the second half of 2009, i.e. the fall wave of the H1N1 (swine flu) pandemic. We find that projected vaccination rates based on sentiments expressed on Twitter are in very good agreement with vaccination rates estimated by the CDC with traditional phone surveys. Looking at the online social network, we find that both negative and positive opinions are clustered, and that an equivalent level of clustering of vaccinations in a population would strongly increase disease outbreak risks.

# INTRODUCTION

Outbreaks of vaccine preventable diseases are a major public health issue. Outbreaks are more likely to occur if either overall vaccination rates decline [1], or if communities with very low vaccination rates increase in frequency or size [2,3]. As individual health behaviors appear to be modulated by social networks [4,5], there is great interest in the dynamics of health behaviors in social networks [6]. Furthermore, measuring health behaviors - such as vaccination - in populations over time and space is essential to identify target areas for interventions and evaluate their effectiveness, but it is generally labor-intensive and expensive when based on traditional survey methodologies [7]. The rise of online social media in the past few years has created new possibilities of measuring health behavior. Such services are used by hundreds of millions of people who are publicly sharing various aspects about their daily lives, including those related to health behavior[8,9]. Using such data to gauge health behaviors in populations represents a fundamental shift in measurement methodology because the study population is not responding to a survey, but rather shares data in a survey-free context, often in real time.

The power of using web data to track events in real time in the context of public health has recently been demonstrated for influenza surveillance [10,11], but assessing health behavior has so far remained elusive. Here, we used publicly available short text messages collected from an online social service (Twitter) from August 2009 to January 2010 in the United States. During this time, pandemic influenza A(H1N1) was spreading nationwide but a vaccine became widely available only very late in the year. We collected practically all publicly available text messages on Twitter (so called "tweets") containing English keywords relating to vaccination as well as location information provided by the authors of text messages (if available). We also collected information on who followed whom among the authors, which allowed us to recreate a directed network of information flow. A subset of the collected tweets was manually evaluated as expressing a negative, positive or neutral sentiment towards influenza A(H1N1) vaccination. We then trained a machine learning algorithm on the manually rated tweets, and then used the resulting classifier to automatically predict sentiments for the remaining unrated text messages. The fully classified data set allowed us to calculate a temporal, localized influenza A(H1N1) vaccination sentiment score and to generate a network of information flow which allowed us to study its properties with respect to the distribution of sentiments. Finally, by extrapolating the findings to empirical contact networks relevant for infectious disease spread, we investigate the effect of non-random vaccination distributions on the likelihood of disease outbreaks.

## RESULTS

Overall, of the 477,768 collected tweets, 318,379 were classified as relevant to the influenza A(H1N1) vacine. Of those, 255,828 were classified as neutral, 26,667 as negative, and 35,884 as positive. Starting from late August 2009, we observed a steady increase in the number of relevant tweets in the United States until early November 2009, after which the number of tweets dropped back to previous levels. Figure 1A shows the absolute numbers of positive ($n_+$), negative ($n_-$) and neutral ($n_0$) tweets per day in the United States. The overall influenza A(H1N1) vaccine sentiment score, measured as the relative difference of positive and negative tweets ( $(n_+ - n_-) / (n_+ + n_- + n_0)$), started at a negative value in late Summer 2009 and showed relatively large short term fluctuations. The 14 day - moving average turned positive in mid October (as the vaccine became available) and remained positive for the rest of the year (Figure 1B).

For vaccination sentiments measured online to be meaningful, they need to be compared to empirical data for validation. A positive correlation between the influenza A(H1N1) vaccination sentiment score and estimated vaccination coverage would be relevant to public health efforts because it would allow for the identification of target areas for communication interventions. To test for such a correlation, we used estimated influenza A(H1N1) vaccination rates up to January 2010 as provided by the CDC [12]. These estimates are based on results from the Behavioral Risk Factor Surveillance System (BRFSS) and the National 2009 H1N1 Flu Survey (NHFS). We found a very strong correlation on the level of HHS regions (weighted r = 0.78, p = 0.017; regions as defined by the US Department of Health & Human Services) using the estimated vaccination coverages for all persons older than 6 months (Figure 1C), and a strong correlation at the level of state (weighted r = 0.52, p = 0.0046). All reported correlation values are Pearson product-moment correlation coefficients because the variables considered for analysis are normally distributed (Shapiro-Wilk test and Anderson-Darling test), weighted by the total number of tweets $(n_+ + n_- + n_0)$ per region.

Using data on who followed whom among users in the dataset allows us to generate a directed network of information flow whose structure (with respect to the distribution of opinions on vaccination) can provide insight into how sentiments are distributed (see Methods). In order to investigate if users preferentially seek information from other users who share their opinion, we measured assortative mixing of users with a qualitatively similar opinion on vaccination (homophily) by calculating the assortativity coefficient *r* which is defined [13] as

$$r = \frac{\sum_i e_{ii} - \sum_i a_i b_i}{1 - \sum_i a_i b_i}$$

where

$$a_i = \sum_j e_{ij} \quad , \quad b_j = \sum_i e_{ij}$$

and $e_{ij}$ is the fraction of edges in the network that connect a node of type $i$ to one of type $j$ (in the direction of $i$ to $j$). A positive value of $r$ (with maximum value 1) is found in a network where nodes are predominantly connected to nodes of the same type. A value of $r = 0$ would indicate a randomly mixed network, and a negative value $\leq -1$ would indicate a disassortative network where nodes of one type are predominantly connected to nodes of the other type (for the technical reasons why the minimum value of $r$ is not always -1 see ref. [13]).

In the network of 39,284 users who had a non-zero sentiment score (from now on referred to as the opinionated network, i.e. containing only users who expressed predominantly either positive or negative opinions), we find $r = 0.144$. In order to assess the significance of this value, we randomized the opinions on the network (bootstrap with replacement) 10,000 times and found the maximum value for $r$ among these randomized networks to be 0.0056, more than an order of magnitude lower than in the original network (mean: $-3*10^{-4}$, 95% CI: -0.0032, 0.0027). We also calculated for each node (user) the fraction $f$ of incoming edges from nodes with the same qualitative sentiment, then randomized the opinions and compared the new distribution of $f$ to the original distribution. For 10,000 randomized networks we found that in all cases the mean of the original distribution (0.601) was significantly larger than the mean of the distribution of the randomized networks ($p < 10^{-95}$ for all tests using paired Wilcoxon signed rank test, max. mean: 0.548, mean of means: 0.531, 95% CI: 0.52, 0.541). These results demonstrate that there is significantly more information flow between users who share the same sentiments than expected based on the distribution of sentiments.

Social networks often naturally divide into communities, i.e groups of people who share common interests, beliefs and opinions. In a network of opinionated users, the question of community structure naturally arises, i.e. are there communities within the network where positive or negative attitudes towards the novel vaccine dominate? In order to tackle this question, we separated the giant component of the opinionated network (34,025 users) into communities of users that are densely connected compared to the rest of the network using the spin glass community detection algorithm [14]. We then calculated the proportion of users with negative attitudes $p(-)$ and compared it to the average in the giant component, $p(-) = 0.396$. With the exception of a single community, all communities (containing at least 1% of the users in the entire network) were significantly more positive or negative than expected (Fig. 2; Fisher's exact test, $10^{-279} < p < 10^{-6}$), ranging from $p(-) = 0.764$ in the most negative community (2,453 users) to $p(-) = 0.266$ in the most positive community (2,517 users).

Non-random distributions of opinions on vaccines can have a profound effect on the likelihood of disease outbreaks if this distribution leads to a clustered distribution of vaccination status in the population [2]. Communities with very low vaccination rates are not protected by herd immunity even if overall vaccination rates in the population are high. To quantify this effect, we used a recently collected high-resolution contact network relevant for infectious disease transmission [15] to simulate the spread if influenza A (H1N1). We performed simulations as described previously [15] with a constant vaccination rate but varying levels of assortativity (see Methods). Fig. 2B shows that the probability of large outbreaks is greatly increased when susceptibility to disease is positively assorted. The probability of an outbreak that infects > 5% of the population, for example, can be increased more than 10-fold at $r > 0.14$ (as observed in the Twitter network) relative to the random distribution where $r \sim 0$ (Figure 2C).

## DISCUSSION

Immunization is generally considered one of the greatest public health achievements in human history [16]. Globally, vaccines have dramatically reduced morbidity and mortality caused by infectious diseases, and vaccines continue to prevent or mitigate the spread of infectious diseases. Despite these resounding successes, however, maintaining sufficiently high vaccination coverages has become very challenging in recent years [1,3]. Unsubstantiated concerns over the safety of vaccines, the rise of the internet and its effect on how fast rumors and misinformation can spread, and a general sense of security from infectious diseases have all contributed to a situation where individual concerns about potential negative side effects often outweigh the benefits of immunization [17]. We've shown here that in a network of almost 40,000 opinionated users of an online social media service, there was significantly more information flow between users who shared the same sentiments than expected if the sentiments were randomly distributed. We also found that most communities were dominated by either positive or negative sentiments towards the novel vaccine. Our data do not allow us to say whether links of information flow were created because of similar vaccination sentiments, or whether other factors, including those that strongly correlate with vaccination sentiments, were responsible for link creation in the first place. Either way, however, the significantly positive assortativity of negative and positive sentiments provide evidence that online social media can act as an "echo chamber" where personal opinions that affect individual medical decisions are predominately reaffirmed by others.

We've also shown that if network clusters of similar sentiments towards vaccination lead to network clusters in the distribution of vaccination, the probability of large outbreaks is greatly increased. Importantly, this effect is strongest when the levels of vaccination coverage are near the levels of required herd immunity under the assumption of a random distribution [2,18]. We do not assume that online social networks strongly overlap with contact networks relevant to infectious disease in the real world, but the extent of homophily may very well be quite similar in both networks (or even lower in online networks such as Twitter where links need not be reciprocated). Indeed, there is increasing evidence from multiple studies that there is significant geographic and social clustering of nonmedical vaccination exemptions, resulting in increases in local risk of vaccine-preventable disease outbreaks [3,19-21]. Communication strategies that aim to decrease the positively assorted distribution of vaccination might thus be an efficient way - in addition to existing efforts to increase general vaccination rates in the population - to decrease the risk of disease outbreaks. Currently, empirical data on assortativity in vaccination status is lacking, but collecting such data would be desirable in order to identify the communities which are at highest risk.

While data from online social media offers great potential to measure health behaviors, and even measure human behavior more generally, a number of caveats need to be mentioned. First, because of the observational nature of the study, we cannot exclude that other confounding factors (e.g. vaccine supply) might have influenced the results. Second, extracting information from short online text messages for the purpose of assessing health behaviors presents a number of challenges. Users of online social media might not be a representative sample of the population; text messages may be interpreted differently by different users, and sentiment analysis is not 100% accurate (see also methods). However, the large volume of data - and the large number of users - substantially reduce the extent to which such limitations affect the overall results. Furthermore, data from online social media have a number of advantages that other data cannot provide. The fact that social media provide network data - i.e. the data do not only contain what is being broadcast, but also to whom - allows us to study processes such as the spread of information, behaviors, opinions, etc. as well as the social structure on which these processes occur. A particular benefit of data provided by Twitter is that they are publicly available - tweets are by default public messages, unlike messages exchanged on other social media services that are generally private by default. This is important because the potential of computational social sciences to understand the dynamics of human societies cannot be fully explored if all data are private and owned by companies and governments[22].

Publicly available data from online social media provide unprecedented opportunities, especially in the realm of public health, e.g. by allowing for inexpensive and efficient tools for the public health community to identify regional areas that would most benefit from intensified communication about the safety and benefit of vaccines. Given the explosive growth of online social media in the past few years, we believe the approach presented here can be applied more generally to study the spread of various health behaviors, a topic of great importance as health behaviors are a leading cause of morbidity and mortality [23].

# METHODS

**Data Collection**

Starting on August 25th 2009, we collected all tweets in English containing at least one of the following search strings: *vaccination* OR *vaccine* OR *vaccinated* OR *vaccinate* OR *vaccinating* OR *immunized* OR *immunize* OR *immunization* OR *immunizing*. Along with the tweet text, we downloaded the date and time when the tweet was published, and the location of the user (if provided). We also downloaded the user id, follower ids, and friends ids. The followers of a user A are those users who will receive messages from user A. The friends of a user A are those users from whom user A receives messages. Thus, information flows from a user to his followers. Until January 19th 2010, we collected tweets every day in real time. Barring occasional short term disruptions due to technical issues, the data set represents the set of tweets meeting the keyword conditions mentioned above in that timeframe.

**Sentiment Analysis**

Each tweet in the dataset needed to be classified into one of four sentiment polarities: *positive*, *negative*, *neutral* and *irrelevant* (see below). Since the dataset of 477,768 tweets was too large to classify manually in a reasonable time frame, we used a machine learning approach to identifying the sentiments expressed in the tweets. This process involved choosing a machine learning algorithm, selecting features of the input and considering other strategies for maximizing the accuracy of the sentiment analysis. We evaluated various classifiers and experimented with different feature sets in order to select the most accurate combination. We compared three standard classification algorithms: Naive Bayes, Maximum Entropy and a

Dynamic Language Model classifier (using process character n-gram models). The Naive Bayes classifier was implemented using the Natural Language Toolkit (NLTK) [24]. The Maximum Entropy classifier was accessed using NLTK, but used the MegaM *(*http://www.cs.utah.edu/~hal/megam/*)* implementation. The Language Model classifier was implemented using LingPipe (http://www.alias-i.com/lingpipe/) and is recommended in the LingPipe documentation for sentiment analysis.

Supervised machine learning approaches, regardless of algorithm used, all require a training dataset. In order to create a training dataset, we needed people to assign sentiment polarities to a random subset of tweets from our database. Study participants ("students" from now on) were recruited from two undergraduate classes at the Pennsylvania State University to rate tweets with the help of a simple web-based rating application. Students were asked to rate tweets based on the following question: what sentiment does the tweeting person (the author) have regarding the influenza A(H1N1) vaccine? They were presented with four options:

1. positive: A positive sentiment means the author is likely to get the influenza A(H1N1) vaccine

    Example tweet that was rated as positive:

    *off to get swine flu vaccinated before work*.

2. negative: A negative sentiment means the author is unlikely to get the influenza A(H1N1) vaccine

    Example tweet that was rated as negative:

*What Can You Do To Resist The U.S. H1N1 "Vaccination" Program? Help Get Word Out. The H1N1 "Vaccine" Is DIRTY.DontGetIt.*

3. neutral: No clear sentiment can be detected

   Example tweet that was rated as neutral:

   *The Health Department will be offering the seasonal flu vaccine for children 6 months - 19 yrs. of age starting on Monday, Nov. 16.*

4. irrelevant: The tweet is not clearly about the influenza A(H1N1) vaccine

   Example tweet that was rated as irrelevant:

   *Filipino discovers new vaccine against malaria that 'treats' the mosquitoes, too!*

The web-based rating application was set up such that every other tweet rated by a student was one that was also rated by all the other students. All other tweets were randomly selected. A student could not rate any tweet more than once, but we did not prevent multiple students from rating the same randomly selected tweet. 64 students volunteered for the task, and submitted 88,237 ratings. Students were assigned to rate at least 1400 tweets, and 44 students complied with this request (the other 20 students rated less than 1400 tweets). In total, students evaluated 47,143 unique tweets.

To evaluate the best classifier, we divided the feature sets extracted from the manually rated tweets into a training set and a test set. The classifiers were evaluated by looking at the percentage of tweets from the test set that classifiers could rate accurately. A tweet was considered accurately rated if the sentiment polarity predicted by the classifier matches majority opinion as assigned by the students. In order to build a high confidence test set, we took all tweets with at least 44 ratings, and then eliminated tweets where both of the following were true: a) the percentage of the majority sentiment polarity was not higher than 50%, i.e. an absolute majority could not be established, **and** b) we (i.e. MS and SK) disagreed with the majority sentiment polarity. This left us with a high confidence test set of 630 tweets. The training set (46,442 tweets) consisted of all the tweets that had less than 44 ratings.

For our sentiment classification, we used an ensemble method combining the Naive Bayes and the Maximum Entropy classifiers. We used the Naive Bayes classifier to determine the positive and negative tweets, and the Maximum Entropy classifier to determine the neutral and irrelevant tweets (leveraging the classifiers' respective strengths). In case of a conflict, the Maximum Entropy classifier's decision was final. The accuracy of this ensemble classifier was 84.29%.

Feature selection is the process of choosing the most informative subset of all the possible features of the training data. Choosing the right features to extract from a tweet is a process of trial and error. We achieved the highest accuracy (for both classifiers) by choosing the set of words constructed from the tweets after filtering out stop words defined by Apache Lucerne's list of stop (http://lucene.apache.org/ - stop words from: org.apache.lucene.analysis.StopAnalyzer) words except "no" and "not". We further filtered out all punctuation except for '!' from the tweet text since exclamation marks are often used to indicate a stronger sentiment. Finally, adding stemming improved the accuracy of the classifier.

The challenge of trying to classify tweets is that each tweet is at maximum only 140 characters. This limitation encourages non-standard abbreviations, slang and otherwise poorly written phrases within the body of a tweet. The general lack of context in a single tweet combined with poorly expressed sentiment means that it is unreasonable to expect a 100% accuracy out of the automated classifiers or even 100% agreement among students. To get a sense of how our automated classifier was performing, we compared it to the accuracy of individual students. Among students that rated at least 1400 tweets, the average accuracy of the students was 64%. Only 7 students had a higher accuracy than the ensemble classifier, the highest being 90%.

To ensure the highest quality during classification of tweets not rated by stundets, we implemented the following strategy. Tweets that were part of the test set didn't need to be evaluated by the classifier since we were confident about the assigned polarities. For all other tweets, we let the classifier predict the sentiment. If the classifier disagreed with the majority, we treated the classifier as a manual rater, and simply took the polarity assigned by the rater (i.e. users and classifier) with the highest accuracy.

**Geocoding**

Twitter allows a tweeter's location to be manually entered into his or her profile. The profile's location field is free form, and accepts any entry; including ones that do not specify any location. A tweeter's position can also be updated by their GPS enabled mobile device, providing an accurate location specified by a pair of latitude and longitude co-ordinates (this feature was not yet widely used in 2009). For this location data to be useful to us, we needed to resolve these location strings into informative locations. Our goal for this study was to resolve within the

United States to the state level. Since we downloaded tweets over a period of several months, there are often multiple tweets from the same tweeter. The location of the tweeter was downloaded with each tweet. In most cases, the tweeter's static location was duplicated multiple times throughout the data set. However, in some cases, over time, tweeters reported different locations. For each tweeter, we resolved all unique locations by using the Yahoo! PlaceFinder web service (http://developer.yahoo.com/geo/placefinder/). Using the PlaceFinder API, we sent the location string to the service which, if it recognized the location, returned state and country level information. Of the 155,676 locations, there were 9,231 location strings that could not be accurately resolved using the PlaceFinder web service. In order to deal with those, we again created a web application and asked students to resolve the location manually if possible. Once all locations were resolved, tweeters with locations in multiple states were excluded from the analysis.

**Network creation**

In order to generate a network of Twitter users that captures information flow about opinions on the influenza A(H1N1) vaccine, we used the following algorithm (recall that information flows from users to their followers):

1. Each user who has at least one relevant (i.e. positive, negative or neutral) tweet is represented by a node in the network.
2. There is a directed edge from user A to user B if at any time point, user B is found among the followers of user A, or user A is found among the friends of user B.

Note that this algorithm treats the network as static, rather than as dynamic. The number of followers and friends might change over time (both generally increase). Thus, the network is essentially a snapshot of the network as seen on the last day of data collection (i.e. January 19th, 2010).

For extremely large counts of followers and friends, the data collection process did only collect the first 5,000 friends and/or followers that Twitter returned - this problem was only recognized late in the project. However, our algorithm deals effectively with this artificial cut off because it searches for edges both in the followers and friends lists. To see why this is the case, assume that a very popular user has more than 5,000 followers. It can be safely assumed that almost all of the followers of this user have a friend count of less than 5,000 users (because accounts with more than 5,000 friends are extremely rare). Thus, even though we might not find an existing edge while looking at the followers of this popular user, we will find the edge while looking at the friends list of the followers. With regard to the cutoff of 5,000 friends (extremely rare), it is unlikely that a person can cognitively follow the message stream of more than 5,000 users, and thus even if we would miss some data, the effect is negligible. Finally, we remove all users from the network that do not have a positive or negative influenza A(H1N1) vaccine sentiment score. The overall influenza A(H1N1) vaccine sentiment score is defined as the difference of positive and negative tweets divided by the sum of all relevant tweets $(n_+ - n_-) / (n_+ + n_- + n_0)$. This results in a network of 39,284 nodes of nodes (i.e. opinionated users) and 685,719 edges (i.e tweets) which has one giant component consisting of 34,025 nodes and 685,390 edges (i.e. 99.95% of the edges of the network are also in the giant component).

**Disease simulations**

We simulate the spread of an influenza A(H1N1)-like infectious disease on an empirical network collected with wireless sensor network technology[15]. We could have chosen any network, but decided to use this network collected at a high school for its high accuracy and coverage, but the results are applicable to any network (indeed we've previously run infectious disease simulations on artificial networks with qualitatively similar results[2]). The disease simulation has been described in detail in [15], but briefly, we use an SEIR simulation model parameterized with data from influenza outbreaks[25,26]. Transmission occurs exclusively along the measured contacts with a minimum duration of 30 minutes throughout the day of the graph collected in ref. [15]. Each individual (i.e., node of the network) is in one of four classes: **s**usceptible, **e**xposed, **i**nfectious, and **r**ecovered. All individuals are initially susceptible with the exception of the vaccinated individuals who are always in the recovered class. On the first day of the week, a random susceptible individual is chosen as the index case, i.e. its status is set to exposed. A simulation is stopped after the number of both exposed and infectious individuals has gone back to 0 (i.e., all infected individuals have recovered). Each time step represents 12 hours and is divided into day and night. Transmission can occur only during the day and only on weekdays (because this is a network collected at a high school and we do not consider any transmission outside of the school; although this assumption will not hold in reality, it allows us to analyze the spread of a disease starting from a single infected case). Transmission of disease from an infectious to a susceptible individual occurs with a probability of $1 - (1 - 0.00767)^w$, where $w$ is the weight of the contact edge (in intervals of 20 seconds - the values were chosen so that the the minimum transmission rate [i.e where $w$=90] is 0.5 - this resulted in an $R_0$ of 2.03 for all outbreaks with at least one secondary infection). After infection, an individual will move into the exposed class (infected but not infectious). After an incubation period, modeled by a right-shifted

Weibull distribution with a fixed offset of half a day [power parameter = 2.21, scale parameter = 1.10 [25]], an exposed individual will become symptomatic and move into the infectious class. As in [15], we assume that on the half day that the individual becomes infectious, the duration of all contacts of the infectious individual is reduced by 75%. This reduction ensures that if an individual becomes symptomatic and starts to feel ill during a school day, social contacts are reduced and the individual leaves the school or is dismissed from school after a few hours. In the following days, all contacts are reduced by 100% until recovery (i.e., the individual stays at home). Once an individual is infectious, recovery occurs with a probability of $1 - 0.95^t$ per time step, where $t$ represents the number of time steps spent in the infectious state [in line with data from an outbreak of influenza A(H1N1) at a New York City school[26]]. After 12 days in the infectious class, an individual will recover if recovery has not occurred before that time. We assume that all exposed individuals developed symptoms (see [15] for a justification of this assumption).

Vaccination occurs by picking individual randomly and vaccinate them to achieve a vaccination coverage of 0.624, reflecting the proportion of positive sentiments in the Twitter network as described above. Such a random process will result in a assortativity index r ~ 0. In order to understand the effect of increased homophily on disease outbreaks, we redistribute the vaccination statuses such that the vaccination coverage remains constant, but $r$ increases. We use a simple algorithm to do to this, given a desired $r$ value:

1. Start with a random distribution of vaccines (i.e. r ~ 0) to achieve a given vaccination coverage.

2. Calculate assortativity index $r$.

3. Pick two random nodes in the network such that one is vaccinated and the other is not.

4. Swap vaccination status of the two nodes and calculate new assortativity index $r^*$.

5. If $r^* \leq r$, undo swap and go to step 3.

6. If $r^* >$ desired value, stop. Otherwise set $r = r^*$ and continue at step 3.

This simple algorithm allows us to use the same network structure with a constant vaccination coverage while at the same time generate variable values of $r$. For each minimum value of $r$, starting from 0 up to 0.145 in increments of 0.005, we generated 100,000 redistributions of vaccinations and ran a simulation per redistribution, as described above. This resulted in a total of 3,000,000 simulation runs on a network with constant structure and constant vaccination coverage, but with different values for the assortativity index $r$.

## ACKNOWLEDGEMENTS

The authors would like to thank A. F. Read, I. M. Cattadori, J. H. Jones and D. R. Hunter for assistance.

# FIGURE LEGENDS

**Fig. 1** (**A**) Total number of negative (red), positive (green), and neutral (blue) tweets relating to influenza A(H1N1) vaccination during the Fall wave of the 2009 pandemic (**B**) Daily (gray) and 14 day moving average (blue) sentiment score during the same time. (**C**) Correlation between estimated vaccination rates for individuals older than 6 months, and sentiment score per HHS region (black dots) and states (gray dots). Numbers represent the ten regions as defined by the US Department of Human Health & Services. Lines shows best fit of linear regression (blue for regions, red for states).

**Fig. 2** (**A**) Proportion of negative sentiments *p(-)* in the network communities. Dashed line shows overall proportion in the opinionated network. The proportions of negative and positive sentiments are significantly different from the overall proportions in the entire opinionated network (with the exception of community E). (**B**) Effect of positive assortativity index (*r*) on relative risk increase (compared to risk at *r*~0) of disease outbreaks that infect at least 3% of the population. Blue line shows best fit of linear regression (confidence interval based on standard error). (**C**) Relative risk increase (compared to risk at *r*~0) of disease outbreaks of a given fraction of the population (on horizontal axis) for two values of assortativity index (*r*), 0.075 (red) and 0.145 (green). Note that the later corresponds to *r* found in the opinionated network (see main text).

**Figure 1**

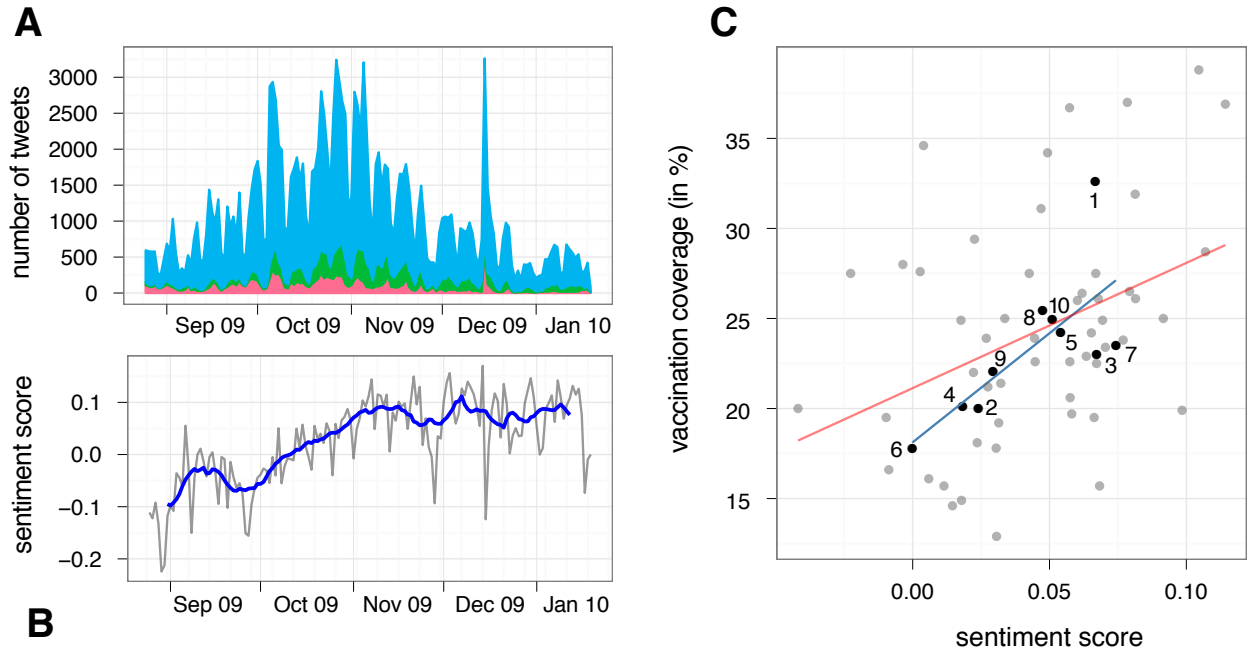

**Figure 2**

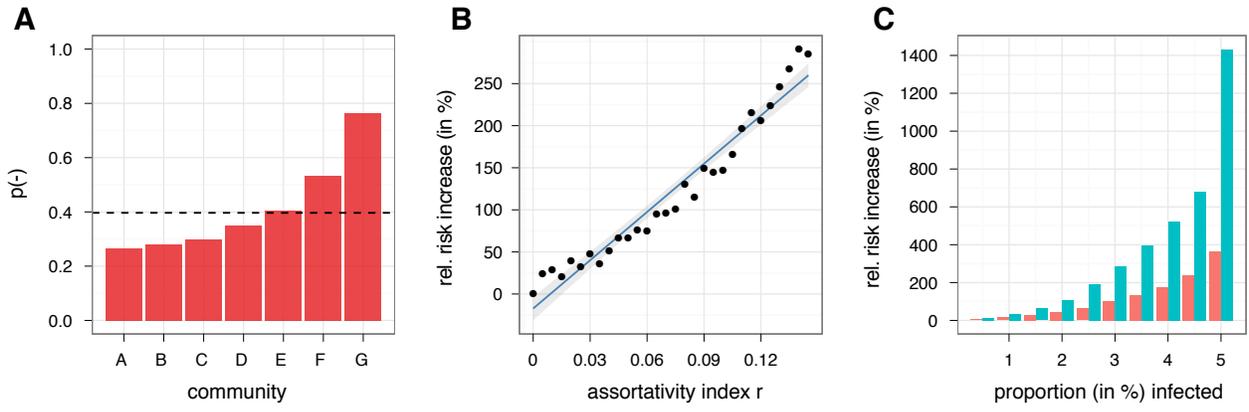